# Prediction of the FIFA World Cup 2018 – A random forest approach with an emphasis on estimated team ability parameters

Andreas Groll [*]   Christophe Ley [†]   Gunther Schauberger [‡]
Hans Van Eetvelde [§]

June 8, 2018

**Abstract** In this work, we compare three different modeling approaches for the scores of soccer matches with regard to their predictive performances based on all matches from the four previous FIFA World Cups 2002 – 2014: *Poisson regression models*, *random forests* and *ranking methods*. While the former two are based on the teams' covariate information, the latter method estimates adequate ability parameters that reflect the current strength of the teams best. Within this comparison the best-performing prediction methods on the training data turn out to be the ranking methods and the random forests. However, we show that by combining the random forest with the team ability parameters from the ranking methods as an additional covariate we can improve the predictive power substantially. Finally, this combination of methods is chosen as the final model and based on its estimates, the FIFA World Cup 2018 is simulated repeatedly and winning probabilities are obtained for all teams. The model slightly favors Spain before the defending champion Germany. Additionally, we provide survival probabilities for all teams and at all tournament stages as well as the most probable tournament outcome.

[*]Statistics Faculty, Technische Universität Dortmund, Vogelpothsweg 87, 44227 Dortmund, Germany, *groll@statistik.tu-dortmund.de*
[†]Faculty of Sciences, Department of Applied Mathematics, Computer Science and Statistics, Ghent University, Krijgslaan 281, 9000 Gent, Belgium, *Christophe.Ley@UGent.be*
[‡]Chair of Epidemiology, Department of Sport and Health Sciences, Technical University of Munich, *g.schauberger@tum.de*
[§]Faculty of Sciences, Department of Applied Mathematics, Computer Science and Statistics, Ghent University, Krijgslaan 281, 9000 Gent, Belgium, *hans.vaneetvelde@ugent.be*

# 1   Introduction

Like the previous FIFA World Cup 2014, also the up-coming tournament in Russia has caught the attention of several modelers who try to predict the tournament winner. One approach that has already produced reasonable results for several of the past European championships (EUROs) and FIFA World Cups is based on the prospective information contained in bookmakers' odds (Leitner, Zeileis, and Hornik, 2010b, Zeileis, Leitner, and Hornik, 2012, 2014, 2016). Nowadays, for such major tournaments bookmakers offer a bet on the winner in advance of the tournament. By aggregating the winning odds from several online bookmakers and transforming those into winning probabilities, inverse tournament simulation can be used to compute team-specific abilities, see Leitner, Zeileis, and Hornik (2010a). With the team-specific abilities all single matches are simulated via paired comparisons and, hence, the complete tournament course is obtained. Using this approach, Zeileis, Leitner, and Hornik (2018) forecast Brazil to win the FIFA World Cup 2018 with a probability of 16.6%, followed by Germany (15.8%) and Spain (12.5%).

The same three teams are determined as the major favorites by a group of experts of the Swiss bank UBS, but with different probabilities and a different order (Audran, Bolliger, Kolb, Mariscal, and Pilloud, 2018): they obtain Germany as top favorite with a winning probability of 24.0%, followed by Brazil (19.8%) and Spain (16.1%). They use a statistical model based on four factors that are supposed to indicate how well a team will be doing during the tournament: the Elo rating, the teams' performances in the qualifications preceding the World Cup, the teams' success in previous World Cup tournaments and a home advantage. The model is calibrated by using the results from the previous five tournaments and 10,000 Monte Carlo simulations are conducted to determine winning probabilities for all teams.

Another model class that has proved of value in predicting the outcome of previous international soccer tournaments, such as EUROs or World Cups, is the class of Poisson regression models which directly model the number of goals scored by both competing teams in the single matches of the tournaments. Let $X_{ij}$ and $Y_{ij}$ denote the goals of the first and second team, respectively, in a match between teams $i$ and $j$, where $i, j \in \{1, \ldots, n\}$ and $n$ denotes the total number of teams in the regarded tournaments. One assumes $X_{ij} \sim Po(\lambda_{ij})$ and $Y_{ij} \sim Po(\mu_{ij})$ where $\lambda_{ij}$ and $\mu_{ij}$ denote the intensity parameters (i.e. the expected number of goals) of the respective Poisson distributions. For these intensity parameters several modeling strategies exist, which incorporate playing abilities or covariates of the competing teams in different ways.

In the simplest case, the Poisson distributions are treated as (conditionally) independent, conditional on the teams' abilities or covariates. For example, Dyte and Clarke (2000) applied this model to data from FIFA world cups and let the Poisson intensities of both competing teams depend on their FIFA ranks. Groll and Abedieh (2013) and Groll, Schauberger, and Tutz (2015) considered a large set of potentially influential variables for EURO and World Cup data, respectively, and used $L_1$-penalized approaches to detect a sparse set of relevant covariates. Based on these, predictions for the EURO 2012 and FIFA World Cup 2014 tournaments were provided. These approaches showed that, when many covariates are regarded and/or the predictive power of the single variables is not clear in advance, regularized estimation approaches can be beneficial.

Many researchers have relaxed the strong assumption of conditional independence and have introduced different possibilities to allow for dependent scores. Dixon and Coles (1997) were the first to identify a (slightly negative) correlation between the scores. As a consequence, they introduced an additional dependence parameter. However, they ignored the fact that the intensity parameters in models including abilities (or covariates) of both teams are themselves correlated. Therefore, even though, conditional on the abilities, the Poisson distributions are assumed to be independent they are marginally correlated. Karlis and Ntzoufras (2003) proposed to model the scores of both teams by a bivariate Poisson distribution, which is able to account for (positive) dependencies between the scores. While the bivariate Poisson distribution can only account for positive dependencies, copula-based models also allow for negative dependencies (see, for example, McHale and Scarf, 2007, McHale and Scarf, 2011 or Boshnakov, Kharrat, and McHale, 2017).

However, with regard to the bivariate Poisson case, Groll, Kneib, Mayr, and Schauberger (2018) provide some evidence that, if highly informative covariates of both competing teams are included into the intensities of both (conditionally) independent Poisson distributions, the dependence structure of the match scores can already be appropriately modeled. They included a large set of covariates for EURO data and used a boosting approach to select a sparse model for the prediction of the EURO 2016. As the dependency parameter of the bivariate Poisson distribution was never updated by the boosting algorithm, two (conditionally) independent Poisson distributions were sufficient.

Closely related to the covariate-based Poisson regression models are Poisson-based ranking methods for soccer teams. The main idea is to find adequate ability parameters that reflect the current strength of the teams best. On basis of a set of matches, those parameters are then estimated by means of maximum likelihood. Ley, Van de Wiele, and Van Eetvelde (2018) have investigated various Poisson models and compared them in terms of their predictive performance. The resulting best models for this purpose are the independent Poisson model and the simplest bivariate Poisson distribution of Karlis and Ntzoufras (2003). Interestingly, Ley et al. (2018) found that those models outperform their competitors both for domestic league matches and national team matches. These statistical strength-based rankings present an interesting alternative to the FIFA ranking.

A fundamentally different modeling approach is based on random (decision) forests – an ensemble learning method for classification, regression and other tasks proposed by Breiman (2001). The method originates from the machine learning and data mining community and operates by first constructing a multitude of so-called decision trees (see, e.g., Quinlan, 1986; Breiman, Friedman, Olshen, and Stone, 1984) on training data. The predictions from the individual trees are then summarized, either by taking the mode of the predicted classes (in classification) or by averaging the predicted values (in regression). This way, random forests reduce the tendency of overfitting and the variance compared to regular decision trees, and, hence, are a common powerful tool for prediction. In preliminary work from Schauberger and Groll (2018) the predictive performance of different types of random forests has been compared on data containing all matches of the FIFA World Cups 2002 – 2014 with conventional regression methods for count data, such as the Poisson models mentioned above. It turned out that random forests provided very satisfactory results and generally outperformed the regression approaches. Moreover, their predictive performances actually were either close to or even outperforming those of the bookmakers, which serve as natural benchmark. These results motivate us to use random forests in the present work to calculate predictions of the up-coming FIFA World Cup 2018. However, we will show that the already excellent predictive power of the random forests can be further increased if adequate estimates of team ability parameters, reflecting the current strength of the national teams, are incorporated as additional covariates.

The rest of the manuscript is structured as follows: in Section 2 we describe the underlying data set covering all matches of the four preceding FIFA World Cups 2002 – 2014. Next, in Section 3 we briefly explain the basic idea of random forests, (regularized) Poisson regression and ranking methods and compare their predictive performances. Then, the best-performing model, which is a combination of random forests and ranking methods, is fitted to the data and used to predict the FIFA World Cup 2018 in Section 4. Finally, we conclude in Section 5.

## 2 Data

In this section, we briefly describe the underlying data set covering all matches of the four preceding FIFA World Cups 2002 – 2014 together with several potential influence variables. Basically, we use the same set of covariates that is introduced in Groll et al. (2015). For each participating team, the covariates are observed either for the year of the respective World Cup (e.g., GDP per capita) or shortly before the start of the World Cup (e.g., FIFA ranking), and, therefore, vary from one World Cup to another.

Several of the variables contain information about the recent performance and sportive success of national teams, as the current form of a national team should have an influence on the team's success in the upcoming tournament. One additional covariate in this regard, which we will introduce later, is reflecting the national teams' current playing abilities. The estimates of these ability parameters are based on a separate Poisson ranking model, see Section 3.3 for details. Beside these sportive variables, also certain economic factors as well as variables describing the structure of a team's squad are collected. We shall now describe in more detail these variables.

**Economic Factors:**

*GDP per capita.* To account for the general increase of the gross domestic product (GDP) during 2002 – 2014, a ratio of the GDP per capita of the respective country and the worldwide average GDP per capita is used (source: http://unstats.un.org/unsd/snaama/dnllist.asp).

*Population.* The population size is used in relation to the respective global population to account for the general world population growth (source: http://data.worldbank.org/indicator/SP.POP.TOTL).

**Sportive factors:**

*ODDSET probability.* We convert bookmaker odds provided by the German state betting agency ODDSET into winning probabilities. The variable hence reflects the probability for each team to win the respective World Cup[1].

*FIFA rank.* The FIFA ranking system ranks all national teams based on their performance over the last four years (source: http://de.fifa.com/worldranking/index.html).

**Home advantage:**

*Host.* A dummy variable indicating if a national team is a hosting country.

*Continent.* A dummy variable indicating if a national team is from the same continent as the host of the World Cup (including the host itself).

*Confederation.* This categorical variable comprises the teams' confederation with six possible values: Africa (CAF); Asia (AFC); Europe (UEFA); North, Central America and Caribbean (CONCACAF); Oceania (OFC); South America (CONMEBOL).

---

[1]The option to bet on the World Champion before the start of the tournament is rather novel. ODDSET, for example, offered the bet for the first time at the FIFA World Cup 2002.

**Factors describing the team's structure:**
> The following variables describe the structure of the teams. They were observed with the 23-player-squad nominated for the respective World Cup.
>
> *(Second) maximum number of teammates.* For each squad, both the maximum and second maximum number of teammates playing together in the same national club are counted.
>
> *Average age.* The average age of each squad is collected.
>
> *Number of Champions League (Europa League) players.* As a measurement of the success of the players on club level, the number of players in the semi finals (taking place only few weeks before the respective World Cup) of the UEFA Champions League (CL) and UEFA Europa League (EL) are counted.
>
> *Number of players abroad/Legionnaires.* For each squad, the number of players playing in clubs abroad (in the season preceding the respective World Cup) is counted.

**Factors describing the team's coach:**
> For the coaches of the teams, *Age* and duration of their *Tenure* are observed. Furthermore, a dummy variable is included, if a coach has the same *Nationality* as his team.

In total, this adds up to 16 variables which were collected separately for each World Cup and each participating team. As an illustration, Table 1 shows the results (1a) and (parts of) the covariates (1b) of the respective teams, exemplarily for the first four matches of the FIFA World Cup 2002. We use this data excerpt to illustrate how the final data set is constructed.

Table 1: Exemplary table showing the results of four matches and parts of the covariates of the involved teams.

| (a) Table of results | | |
|---|---|---|
| FRA  | 0:1 |  SEN |
| URU  | 1:2 |  DEN |
| FRA  | 0:0 |  URU |
| DEN  | 1:1 |  SEN |
| ⋮ | ⋮ | ⋮ |

| (b) Table of covariates | | | | | |
|---|---|---|---|---|---|
| World Cup | Team | Age | Rank | Oddset | … |
| 2002 | France | 28.3 | 1 | 0.149 | … |
| 2002 | Uruguay | 25.3 | 24 | 0.009 | … |
| 2002 | Denmark | 27.4 | 20 | 0.012 | … |
| 2002 | Senegal | 24.3 | 42 | 0.006 | … |
| ⋮ | ⋮ | ⋮ | ⋮ | ⋮ | ⋱ |

For the modeling techniques that we shall introduce in the following sections, all of the metric covariates are incorporated in the form of differences. For example, the final variable *Rank* will be the difference between the FIFA ranks of both teams. The categorical variables *Host*, *Continent*, *Confederation* and *Nationality*, however, are included as separate variables for both competing teams. For the variable *Confederation*, for example, this results in two columns of the corresponding design matrix denoted by *Confed* and *Confed.Oppo*, where *Confed* is referring to the confederation of the first-named team and *Confed.Oppo* to the one of its opponent.

As we use the number of goals of each team directly as the response variable, each match corresponds to two different observations, one per team. For the covariates, we consider differences which are computed from the perspective of the first-named team. For illustration, the resulting final data structure for the exemplary matches from Table 1 is displayed in Table 2.

Table 2: Exemplary table illustrating the data structure.

| Goals | Team    | Opponent | Age   | Rank | Oddset | ... |
|------:|---------|----------|------:|-----:|-------:|-----|
| 0     | France  | Senegal  | 4.00  | -41  | 0.14   | ... |
| 1     | Senegal | France   | -4.00 | 41   | -0.14  | ... |
| 1     | Uruguay | Denmark  | -2.10 | 4    | -0.00  | ... |
| 2     | Denmark | Uruguay  | 2.10  | -4   | 0.00   | ... |
| 0     | France  | Uruguay  | 3.00  | -23  | 0.14   | ... |
| 0     | Uruguay | France   | -3.00 | 23   | -0.14  | ... |
| 1     | Denmark | Senegal  | 3.10  | -22  | 0.01   | ... |
| 1     | Senegal | Denmark  | -3.10 | 22   | -0.01  | ... |
| ⋮     | ⋮       | ⋮        | ⋮     | ⋮    | ⋮      | ⋱   |

Note that in our final model used in Section 4 for the prediction of the FIFA World Cup 2018 tournament, we incorporate another covariate, namely estimates of the teams' playing ability parameters, which are based on a separate Poisson ranking model (see Section 3.3 for details).

# 3 Methods

In this section, we briefly describe several different methods that generally come into consideration when the goals scored in single soccer matches are directly modeled. Actually, all of them (or slight modifications thereof) have already been used in former research on soccer data and, generally, all yielded satisfactory results. However, we aim to choose the approach that has the best performance regarding prediction and then use it to predict the FIFA World Cup 2018.

## 3.1 Random forests

Random forests were originally proposed by Breiman (2001) and are nowadays seen as a mixture between statistical modeling and machine learning. They are an aggregation of a (large) number of classification or regression trees (CARTs). CARTs (Breiman et al., 1984) repeatedly partition the predictor space (mostly using binary splits). The goal of the partitioning process is to find partitions such that the respective response values are very homogeneous within a partition but very heterogeneous between partitions. CARTs can be used both for metric response (regression trees) and for nominal/ordinal responses (classification trees). The most frequent visualization tool for CARTs is the so-called dendrogram (see also Figure 1). For prediction, all response values within a partition are aggregated either by averaging (in regression trees) or simply by counting and using majority vote (in classification trees).

In this work, we will try to use trees (and, accordingly, random forests) for the prediction of the number of goals a team scores in a match of a FIFA World Cup. For that purpose, we use the predictor variables introduced in Section 2. As an illustrating example, Figure 1 shows the dendrogram for a regression tree applied to these data using the function `ctree` from the R-package `party` (Hothorn, Bühlmann, Dudoit, Molinaro, and van der Laan, 2006).

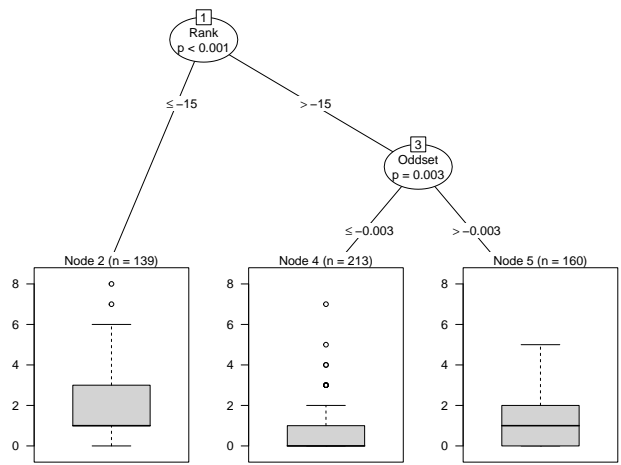

Figure 1: Exemplary regression tree for FIFA World Cup 2002 – 2014 data. Number of goals is used as response variable, variables described in Section 2 are used as predictors.

Only two splits are performed in this example, one for the variable *Rank* and one for *Oddset*, which leads to a total of 3 partitions in the predictor space. The boxplots corresponding to each of the 3 final partitions show the distribution of the response (number of goals) for all observations falling into the respective nodes. In principle, one could of course perform many more splits, finally leading to perfectly separated partitions where each partition only contains observations referring to the same value for the response variable. However, typical regression trees are "pruned" to prevent overfitting to the training data.

As mentioned before, random forests are the aggregation of a large number of trees. The combination of many trees has the advantage that the resulting predictions inherit the feature of unbiasedness from the single trees while reducing the variance of the predictions. The single trees are grown independently from each other. To get a final prediction, predictions of single trees are aggregated, in our case of regression trees simply by averaging over all the predictions from the single trees. In order to achieve the goal that the aggregation of trees is less variant than a single tree, it is important to reduce the dependencies between the trees that are aggregated to a forest. Typically, two randomisation steps are applied to achieve this goal. First, the trees are not applied to the original sample but to bootstrap samples or random subsamples of the data. Second, at each node a (random) subset of the predictor variables is drawn which are used to find the best split. These steps de-correlate the single trees and help to lower the variance of a random forest compared to single trees.

In contrast to regression trees, random forests are much harder to visualize and to interpret. While in trees the effect of a single predictor can (almost) be seen at one glance when looking at the respective dendrogram, this is almost impossible for random forests. Each predictor may have different effects (or no effect at all) in different trees. The best way to nevertheless understand the role of the single predictor variables is the so-called variable importance. Typically, the variable importance of a predictor is measured by permuting each of the predictors separately in the out-of-bag observations of each tree. Out-of-bag observations are observations which are not part of the respective subsample or bootstrap sample that is used to fit a tree. Permuting a variable means that within the variable each value is randomly assigned to a location within the vector. If, for example, Age is permuted, the average age of the German team in 2002 could be assigned to the average age of the Brazilian team in 2010. When permuting variables randomly, they lose their information with respect to the response variable (if they have any). Then, one measures the loss of prediction accuracy compared to the case where the variable is not permuted. Permuting variables with a high importance will lead to a higher loss of prediction accuracy than permuting values with low importance. For the sake of illustration, Figure 2 shows bar plots for a random forest applied to the World Cup data introduced in Section 2.

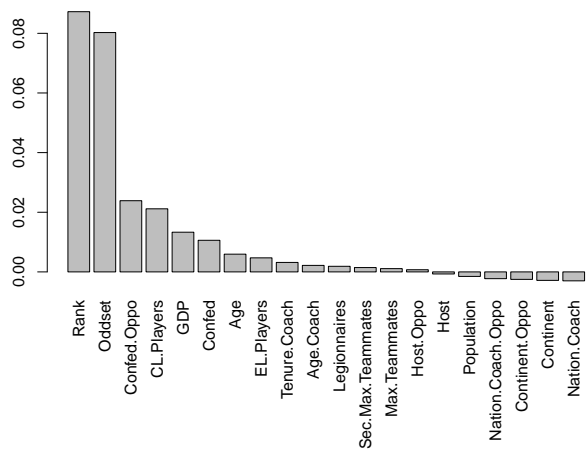

Figure 2: Bar plot displaying the variable importance in a random forest applied to FIFA World Cup 2002 – 2014 data. Number of goals is used as response variable, variables described in Section 2 are used as predictors.

It can be seen that the most important predictors are Rank, Oddset, CL.Players and Confed.Oppo. This finding is in line with the findings in Groll et al. (2015) in the context of Lasso estimation.

In R (R Core Team, 2018), two slightly different variants of regression forests are available. First, the classical random forest algorithm proposed by Breiman (2001) from the R-package ranger (Wright and Ziegler, 2017). The second variant is implemented in the function cforest from the party package. Here, the single trees are constructed following the principle of conditional inference trees as proposed in Hothorn et al. (2006). The main advantage of these conditional inference trees is that they avoid selection bias in cases where the covariates have different scales, e.g., numerical vs. categorical with many categories (see, for example, Strobl, Boulesteix, Zeileis, and Hothorn, 2007, and Strobl, Boulesteix, Kneib, Augustin, and Zeileis, 2008, for details). Conditional forests share the feature of conditional inference trees of avoiding biased variable selection. Additionally, single predictions are not simply averaged but observation weights are used (see Hothorn, Lausen, Benner, and Radespiel-Tröger, 2004).

For prediction, the basic principle is that a predefined number of trees $B$ (e.g., $B = 5000$) is fitted to (bootstrap samples of) the training data. To predict a new observation, its covariate values are dropped down each of the regression trees, resulting in $B$ predictions. The average of those is then used as a point estimate of the expected numbers of goals conditioning on the covariate values. However, these point estimates cannot directly be used for the prediction of the outcome of single matches or a whole tournament. First of all, plugging in both predictions corresponding to one match does not necessarily deliver an integer outcome (i.e., a result). For example, one might get predictions of 2.3 goals for the first and 1.1 goals for the second team. Furthermore, as no explicit distribution is assumed for these predictions it is not possible to randomly draw results for the respective match. Hence, similar to the regression methods described in the next section, we will treat the predicted expected value for the number of goals as an estimate for the intensity $\lambda$ of a Poisson distribution $Po(\lambda)$. This way we can randomly draw results for single matches and compute probabilities for the match outcomes *win*, *draw* and *loss* by using two independent Poisson distributions (conditional on the covariates) for both scores.

Besides regression forests modeling the exact number of goals, principally also random forests for the categorial (ordinal) match outcome *win*, *draw* and *loss* can be applied. Though, obviously, these forests cannot directly be used for the simulation of exact match outcomes, Schauberger and Groll (2018) explain how to suitably combine them with a random forest predicting the number of goals.

Altogether, in the preliminary work from Schauberger and Groll (2018) the predictive performance of these different random forest approaches has been compared and it turned out that the `cforest` from the `party` package yielded the best results. For this reason, in the remainder of this work we will focus on this specific approach (from now on simply referred to as *Random Forest*).

## 3.2 Regression

An alternative, more traditional approach which is often applied for modeling soccer results is based on regression. In the most popular case the scores of the competing teams are treated as (conditionally) independent variables following a Poisson distribution (conditioned on certain covariates), as introduced in the seminal works of Maher, 1982 and Dixon and Coles, 1997. Similar to the random forests from the previous section, the methods described in this section can also be directly applied to data in the format of Table 2 from Section 2. Hence, each score is treated as a single observation and one obtains two observations per match. Accordingly, for *n* teams the respective model has the form

$$\begin{aligned}
Y_{ijk}|x_{ik},x_{jk} &\sim Po(\lambda_{ijk}), \\
\log(\lambda_{ijk}) &= \beta_0 + (x_{ik}-x_{jk})^\top \beta + z_{ik}^\top \gamma + z_{jk}^\top \delta,
\end{aligned} \quad (1)$$

where $Y_{ijk}$ denotes the score of team $i$ against team $j$ in tournament $k$ with $i,j \in \{1,\ldots,n\}$, $i \neq j$. The metric characteristics of both competing teams are captured in the $p$-dimensional vectors $x_{ik}, x_{jk}$, while $z_{ik}$ and $z_{jk}$ capture dummy variables for the categorical covariates *Host*, *Continent*, *Confed* and *Nation.Coach* (built, for example, by reference encoding), separately for the considered teams and their respective opponents. For these variables, it is not sensible to build differences between the respective values. Furthermore, $\beta$ is a parameter vector which captures the linear effects of all metric covariate differences and $\gamma$ and $\delta$ collect the effects of the dummy variables corresponding to the teams and their opponents, respectively. For notational convenience, we collect all covariate effects in the $\tilde{p}$-dimensional real-valued vector $\theta^\top = (\beta^\top, \gamma^\top, \delta^\top)$.

If, as in our case, several covariates of the competing teams are included into the model it is sensible to use regularization techniques when estimating the models to allow for variable selection and to avoid overfitting. In the following, we will introduce such a basic regularization approach, namely the conventional Lasso (Tibshirani, 1996). For estimation, instead of the regular likelihood $l(\beta_0, \theta)$ the penalized likelihood

$$l_p(\beta_0, \theta) = l(\beta_0, \theta) + \lambda P(\beta_0, \theta) \qquad (2)$$

is maximized, where $P(\beta_0, \theta) = \sum_{v=1}^{\tilde{p}} |\theta_v|$ denotes the ordinary Lasso penalty with tuning parameter $\lambda$. The optimal value for the tuning parameter $\lambda$ will be determined by 10-fold cross-validation (CV). The model will be fitted using the function `cv.glmnet` from the R-package `glmnet` (Friedman, Hastie, and Tibshirani, 2010). In contrast to the similar ridge penalty (Hoerl and Kennard, 1970), which penalizes squared parameters instead of absolute values, Lasso does not only shrink parameters towards zero, but is able to set them to exactly zero. Therefore, depending on the chosen value of the tuning parameter, Lasso also enforces variable selection.

As a possible extension of the model (1), the linear predictor can be augmented by team-specific attack and defense effects for all competing teams. This extension was used in Groll et al. (2015) to predict the FIFA World Cup 2014. There, the two effects corresponding to the same team have been treated as a group of parameters and, hence, the Group Lasso penalty proposed by (Yuan and Lin, 2006) has been applied on those parameter groups.

Besides the above mentioned approaches, there is a variety of alternative regularization methods available. For example, if the model (1) shall be extended from linear to smooth covariate effects $f(\cdot)$ for metric covariates, boosting techniques designed for generalized additive models could be used, such as the `gamboost` algorithm from the `mboost` package (Hothorn, Buehlmann, Kneib, Schmid, and Hofner, 2017). Principally, when considering distributions for count data, alternatively to the Poisson distribution the negative binomial distribution could be used as response distribution, which is less restrictive, as it overcomes the rather strict assumption of the expectation equating the variance. In Schauberger and Groll (2018) two different boosting approaches for this model class have been investigated. However, no overdispersion compared to the Poisson assumption was detected and, hence, the models reduced back to the Poisson case.

Altogether, in Schauberger and Groll (2018) the simple Lasso from (2) with predictor structure (1) turned out to be the best-performing regression approach, though being slightly outperformed by all random forests from Section 3.1. Hence, for the remainder of this work we will concentrate on the conventional Lasso (from now on simply referred to as *Lasso*).

## 3.3 Ranking methods

In this section we describe how Poisson models can be used to lead to rankings that reflect a team's current ability. We will restrict our attention to the two best-performing models according to the comparison achieved in Ley et al. (2018). The main idea consists in assigning a strength parameter to every team and in estimating those parameters over a period of $M$ matches via weighted maximum likelihood, where the weights are of two types: time depreciation and match importance.

We start by describing the two common traits underpinning both rankings, namely the weights. The time decay function is defined as follows: a match played $x_m$ days back gets a weight of

$$w_{time,m}(x_m) = \left(\frac{1}{2}\right)^{\frac{x_m}{\text{Half period}}}, \qquad (3)$$

meaning that a match played *Half period* days ago only contributes half as much as a match played today and a match played $3\times$*Half period* days ago contributes 12.5% of a match played today. This ensures that recent matches receive more importance and leads to the desired current-strength ranking. The match importance weights are directly inherited from the official FIFA ranking and can take the values 1 for a friendly game, 2.5 for a confederation or world cup qualifier, 3 for a confederation tournament (e.g. UEFA EUROs or the Africa Cup of Nations) or the confederations cup, and 4 for World Cup matches. The relative importance of a national match is indicated by $w_{type,m}$ for $m = 1,\ldots,M$.

The independent Poisson ranking model looks very similar to the Poisson regression model (1) described above. If we have $M$ matches featuring a total of $n$ teams, we write

$$\begin{aligned} Y_{ijm} &\sim Po(\lambda_{ijm}), \\ \log(\lambda_{ijm}) &= \beta_0 + (r_i - r_j) + h \cdot \mathbb{I}(\text{team } i \text{ playing at home}), \end{aligned} \qquad (4)$$

where now $Y_{ijm}$ stands for the number of goals scored by team $i$ against team $j$ ($i, j \in \{1, ..., n\}$) in match $m$ (where $m \in \{1, ..., M\}$), $\lambda_{ijm}$ is the expected number of goals for team $i$ in this match and $r_i$ and $r_j$ are the strengths of team $i$ and team $j$. The last term $h$ represents the home effect and is only added if team $i$ plays at home. This leads to the likelihood function

$$L = \prod_{m=1}^{M} \left( \frac{\lambda_{ijm}^{y_{ijm}}}{y_{ijm}!} \exp(-\lambda_{ijm}) \cdot \frac{\lambda_{jim}^{y_{jim}}}{y_{jim}!} \exp(-\lambda_{jim}) \right)^{w_{type,m} \cdot w_{time,m}}, \quad (5)$$

where $y_{ijm}$ and $y_{jim}$ stand for the actual number of goals scored by teams $i$ and $j$ in match $m$. The values of the strength parameters $r_1, \ldots, r_n$, which determine the resulting ranking, are computed numerically as maximum likelihood estimates.

A covariance parameter $\lambda_{C_m}, m = 1, \ldots, M$, is added for the bivariate Poisson model of Karlis and Ntzoufras (2003). The joint probability function of the home and away score is then given by the bivariate Poisson probability mass function (pmf) with parameters $\lambda_{ijm}$, $\lambda_{jim}$ and $\lambda_{C_m}$:

$$P(Y_{ijm} = z, Y_{jim} = y) = \frac{\lambda_{ijm}^z \lambda_{jim}^y}{z! y!} \exp(-(\lambda_{ijm} + \lambda_{jim} + \lambda_{C_m})) \sum_{k=0}^{\min(z,y)} \binom{z}{k}\binom{y}{k} k! \left( \frac{\lambda_{C_i}}{\lambda_{ijm}\lambda_{jim}} \right)^k, \quad (6)$$

where $Y_{ijm}$ and $Y_{jim}$, respectively, stand for the random variables "goals scored by teams $i$ and $j$ in match $m$". The parameters $\lambda_{ijm}$ and $\lambda_{jim}$ are defined in the same way as for the independent Poisson model and, hence, incorporate a team's strength or ability parameter. There exist various proposals for defining $\lambda_{C_m}$; quite conveniently, the best choice according to the comparison study of Ley et al. (2018) corresponds to a constant covariance parameter $\lambda_{C_m} = \beta_C$. Clearly, we retrieve the independent Poisson when $\lambda_{C_m} = 0$. The likelihood function is of the form (5) with the bivariate pmf instead of the independent version, and the maximization is achieved numerically.

We refer the interested reader to Section 2 of Ley et al. (2018) for more variants of the bivariate Poisson model. We also note that, instead of a time depreciation effect, time series models where the strength parameters vary in time could have been used (Koopman and Lit, 2015).

While Ley et al. (2018) only considered the matches of the European teams, we include all international team matches here. Team strengths are estimated by taking into account all matches in the previous 8 years. We selected the best model and the best Half Period parameter based on the predictive performance of the models on the international soccer data from 2002 to 2017. As can be seen in Table 3, the Bivariate Poisson model with a Half Period of 3 years is selected as the best ranking model according to the average Rank Probability Score (RPS; Gneiting and Raftery, 2007), which is defined in the next section. From now on this model is simply referred to as *Ranking*.

Table 3: Comparison of the predictive performance of the independent and bivariate Poisson models with Half period values of about 1,2,3,4 and 5 years. The best model is the model with the lowest average RPS.

|    | method              | Half Period | average RPS |
|----|---------------------|-------------|-------------|
| 1  | Bivariate Poisson   | 1095        | 0.17366     |
| 2  | Bivariate Poisson   | 730         | 0.17369     |
| 3  | Bivariate Poisson   | 1460        | 0.17382     |
| 4  | Independent Poisson | 1095        | 0.17382     |
| 5  | Independent Poisson | 730         | 0.17384     |
| 6  | Independent Poisson | 1460        | 0.17397     |
| 7  | Bivariate Poisson   | 1825        | 0.17398     |
| 8  | Independent Poisson | 1825        | 0.17415     |
| 9  | Bivariate Poisson   | 365         | 0.17555     |
| 10 | Independent Poisson | 365         | 0.17565     |

## 3.4 Combining methods

The three different approaches introduced in Sections 3.1 - 3.3 are now compared with regard to their predictive performance. For this purpose, we apply the following general procedure on the World Cup 2002 – 2014 data:

1. *Form a training data set containing three out of four World Cups.*
2. *Fit each of the methods to the training data.*
3. *Predict the left-out World Cup using each of the prediction methods.*
4. *Iterate steps 1-3 such that each World Cup is once the left-out one.*
5. *Compare predicted and real outcomes for all prediction methods.*

This procedure ensures that each match from the total data set is once part of the test data and we obtain out-of-sample predictions for all matches. In step *5*, several different performance measures for the quality of the predictions are investigated.

Let $\tilde{y}_i \in \{1,2,3\}$ be the true ordinal match outcomes for all $i = 1,\ldots,N$ matches from the four considered World Cups. Additionally, let $\hat{\pi}_{1i}, \hat{\pi}_{2i}, \hat{\pi}_{3i}$, $i = 1,\ldots,N$, be the predicted probabilities for the match outcomes obtained by one of the different methods introduced in Sections 3.1 - 3.3. These can be computed by assuming that the numbers of goals follow (conditionally) independent Poisson distributions, where the event rates $\lambda_{1i}$ and $\lambda_{2i}$ for the scores of match $i$ are estimated by the respective predicted expected values. Let $G_{1i}$ and $G_{2i}$ denote the random variables representing the number of goals scored by two competing teams in match $i$. Then, the probabilities $\hat{\pi}_{1i} = P(G_{1i} > G_{2i}), \hat{\pi}_{2i} = P(G_{1i} = G_{2i})$ and $\hat{\pi}_{3i} = P(G_{1i} < G_{2i})$, which are based on the corresponding Poisson distributions $G_{1i} \sim Po(\hat{\lambda}_{1i})$ and $G_{2i} \sim Po(\hat{\lambda}_{2i})$ with estimates $\hat{\lambda}_{1i}$ and $\hat{\lambda}_{2i}$, can be easily calculated via the Skellam distribution. Based on these predicted probabilities, we use three different performance measures to compare the predictive power of the methods:

- the multinomial *likelihood*, which for a single match outcome is defined as $\hat{\pi}_{1i}^{\delta_{1\tilde{y}_i}} \hat{\pi}_{2i}^{\delta_{2\tilde{y}_i}} \hat{\pi}_{3i}^{\delta_{3\tilde{y}_i}}$, with $\delta_{r\tilde{y}_i}$ denoting Kronecker's delta. It reflects the probability of a correct prediction. Hence, a large value reflects a good fit.
- the *classification rate*, based on the indicator functions $\mathbb{I}(\tilde{y}_i = \arg\max_{r \in \{1,2,3\}} (\hat{\pi}_{ri}))$, indicating whether match $i$ was correctly classified. Again, a large value of the classification rate reflects a good fit.
- the *rank probability score* (RPS) which, in contrast to both measures introduced above, explicitly accounts for the ordinal structure of the responses. For our purpose, it can be defined as $\frac{1}{3-1} \sum_{r=1}^{3-1} \left( \sum_{l=1}^{r} (\hat{\pi}_{li} - \delta_{l\tilde{y}_i}) \right)^2$. As the RPS is an error measure, here a low value represents a good fit.

Odds provided by bookmakers serve as a natural benchmark for these predictive performance measures. For this purpose, we collected the so-called "three-way" odds for (almost) all matches of the FIFA World Cups 2002 – 2014[2]. By taking the three quantities $\tilde{\pi}_{ri} = 1/\text{odds}_{ri}, r \in \{1,2,3\}$, of a match $i$ and by normalizing with $c_i := \sum_{r=1}^{3} \tilde{\pi}_{ri}$ in order to adjust for the bookmaker's margins, the odds can be directly transformed into probabilities using $\hat{\pi}_{ri} = \tilde{\pi}_{ri}/c_i$ [3].

Table 4 displays the results for these (ordinal) performance measures for the methods introduced in Sections 3.1 - 3.3 as well as for the bookmakers, averaged over 250 matches from the four FIFA World Cups 2002 – 2014. It turns out that the ranking method yields very good results with respect to multinomial likelihood, while the random forest method performs very well in terms of the classification rate. Both methods come close to or even outperform the bookmakers for these criteria. The Lasso method also yields satisfactory results with respect to most criteria, only in terms of RPS it is clearly outperformed by the other methods.

---

[2]Three-way odds consider only the match tendency with possible results *victory team 1*, *draw* or *defeat team 1* and are usually fixed some days before the corresponding match takes place. The three-way odds were obtained from the website http://www.betexplorer.com/. Unfortunately, for 6 matches from the FIFA World Cup 2006 no odds were available and, hence, the results from Table 4 are based on 250 matches only.

[3]The transformed probabilities implicitly assume that the bookmaker's margins are equally distributed on the three possible match tendencies.

Table 4: Comparison of the prediction methods for ordinal match outcomes.

|               | Likelihood | Class. Rate | RPS   |
|---------------|------------|-------------|-------|
| Random Forest | 0.410      | 0.548       | 0.192 |
| Lasso         | 0.419      | 0.524       | 0.198 |
| Ranking       | 0.415      | 0.532       | 0.190 |
| Bookmakers    | 0.425      | 0.524       | 0.188 |

As we later want to predict both winning probabilities for all teams and the whole tournament course for the FIFA World Cup 2018, we are also interested in the performance of the regarded methods with respect to the prediction of the exact number of goals. In order to identify the teams that qualify for the knockout stage, the precise final group standings need to be determined. To be able to do so, the precise results of the matches in the group stage play a crucial role[4].

For this reason, we also evaluate the methods' performances with regard to the quadratic error between the observed and predicted number of goals for each match and each team, as well as between the observed and predicted goal difference. Let now $y_{ijk}$, for $i, j = 1, \ldots, n$ and $k \in \{2002, 2006, 2010, 2014\}$, denote the observed numbers of goals scored by team $i$ against team $j$ in tournament $k$ and $\hat{y}_{ijk}$ a corresponding predicted value, obtained by one of the methods from 3.1 - 3.3. Then we calculate the two quadratic errors $(y_{ijk} - \hat{y}_{ijk})^2$ and $\big((y_{ijk} - y_{jik}) - (\hat{y}_{ijk} - \hat{y}_{jik})\big)^2$ for all $N$ matches of the four FIFA World Cups 2002 – 2014. Finally, per method we calculate (mean) quadratic errors. Note that in this case the odds provided by the bookmakers cannot be used for comparison. So, in contrast to Table 4 where six matches had to be left out due to missing bookmaker information, now all $N = 256$ matches are used. Table 5 shows that the ranking method and the random forest perform comparably well while Lasso performs worse than its competitors.

---

[4]The final group standings are determined by (1) the number of points, (2) the goal difference and (3) the number of scored goals. If several teams coincide with respect to all of these three criteria, a separate chart is calculated based on the matches between the coinciding teams only. Here, again the final standing of the teams is determined following criteria (1)–(3). If still no distinct decision can be taken, the decision is induced by lot.

Table 5: Comparison of the prediction methods for the exact number of goals and the goal difference based on mean quadratic error.

|  | Goal Difference | Goals |
|---|---:|---:|
| Random Forest | 2.543 | 1.330 |
| Lasso | 2.835 | 1.421 |
| Ranking | 2.560 | 1.349 |

Now, the team abilities $r_i$ and $r_j$ introduced in equation (4) (though they actually stem from the bivariate Poisson model, which was best-performing in Table 3) can principally also be seen as another covariate (similar to and, in fact, even more informative than e.g. the FIFA ranking). For this reason, we decided to include them both in the random forest and the Lasso method as an additional covariate. And, indeed, it turned out that both methods clearly improved with regard to all five criteria, see Table 6 and 7 (compared to Table 4 and 5). Especially the random forest method now yields very satisfactory results with regard to all (ordinal) performance measures (see Table 6), being even better than the bookmakers for two criteria. These ordinal performance measures are of particular importance for a good tournament prediction, as they are closely related to correct predictions of single matches. Additionally, also the mean quadratic errors of the exact number of goals and of the goal difference have clearly improved and the random forest method now outperforms both competitors in these criteria. Altogether, based on these results we assess the random forest method combined with the team abilities as an additional covariate to be the most promising method for the prediction of the FIFA World Cup 2018 tournament. Hence, the predictions in the next section are based on this approach.

Table 6: Comparison of the different prediction methods for ordinal match outcomes with abilities included as covariates.

|  | Likelihood | Class. Rate | RPS |
|---|---:|---:|---:|
| Random Forest | 0.419 | 0.556 | 0.187 |
| Lasso | 0.429 | 0.540 | 0.194 |
| Ranking | 0.415 | 0.532 | 0.190 |
| Bookmakers | 0.425 | 0.524 | 0.188 |

Table 7: Comparison of different prediction methods for the exact number of goals and the goal difference based on mean quadratic error with abilities included as covariates.

|  | Goal Difference | Goals |
|---|---:|---:|
| Random Forest | 2.473 | 1.296 |
| Lasso | 2.809 | 1.427 |
| Ranking | 2.560 | 1.349 |

To conclude this section, we fit the random forest approach including the team abilities as an additional predictor variable to the complete data set covering all World Cups from 2002 to 2014. Figure 3 shows the corresponding bar plots for the variable importance of the single predictors. Interestingly, the abilities are by far the most important predictor in the random forest and carry clearly more information than all other predictors (see also Figure 2). In the following section, this model will be applied to (new) data for the upcoming World Cup 2018 in Russia to predict winning probabilities for all teams and to predict the tournament course.

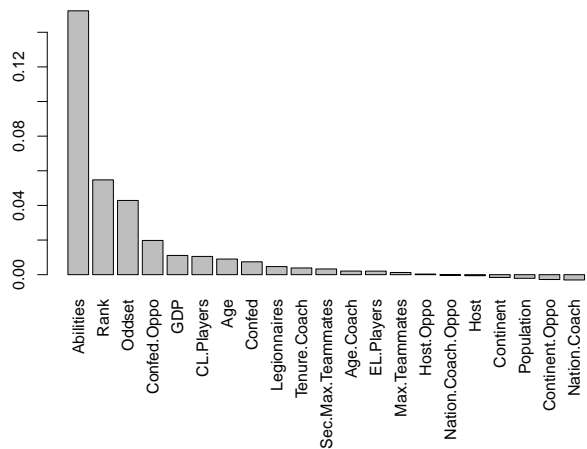

Figure 3: Bar plot displaying the variable importance in a random forest applied to FIFA World Cup data when ability estimates are used as predictors additionally to the variables described in Section 2.

# 4 Prediction of the FIFA World Cup 2018

In this section we apply the best-performing model from Section 3.4, namely the combination of a random forest with adequate team ability estimates from a ranking method, to the World Cup 2018 data. The abilities were estimated by the bivariate Poisson model with a half period of 3 years. All matches of the 228 national teams played since 2010-06-13 up to 2018-06-06 are used for the estimation, what results in a total of more than 7000 matches. All further predictor variables are taken as the latest values shortly before the World Cup (and using the finally announced squads of 23 players for all nations).

## 4.1 Probabilities for FIFA World Cup 2018 Winner

For each match in the World Cup 2018, the random forest can be used to predict an expected number of goals for both teams. Given the expected number of goals, a real result is drawn by assuming two (conditionally) independent Poisson distributions for both scores. Based on these results, all 48 matches from the group stage can be simulated and final group standings can be calculated. Due to the fact that real results are simulated, we can precisely follow the official FIFA rules when determining the final group standings[5]. This enables us to determine the matches in the round-of-sixteen and we can continue by simulating the knockout stage. In the case of draws in the knockout stage, we simulate extra-time by a second simulated result. However, here we multiply the expected number of goals by the factor 0.33 to account for the shorter time to score (30 min instead of 90 min). In the case of a further draw in extra-time we simulate the penalty shootout by a (virtual) coin flip.

Following this strategy, a whole tournament run can be simulated, which we repeat 100,000 times. Based on these simulations, for each of the 32 participating teams probabilities to reach the single knockout stages and, finally, to win the tournament are obtained. These are summarized in Table 8 together with the winning probabilities based on the ODDSET odds for comparison.

We can see that, according to our random forest model, Spain is the favored team with a predicted winning probability of 17.8% followed by Germany, Brazil, France and Belgium. Overall, this result seems in line with the probabilities from the bookmakers, as we can see in the last column. While Oddset favors Germany and Brazil, the random forest model predicts a slight advantage for Spain. However, we can see no clear favorite, several teams seem to have good chances. Besides the probabilities of becoming world champion, Table 8 provides some further interesting insights also for the single stages within the tournament. For example, it is interesting to see that the two favored teams Spain and Germany have almost equal chances to at least reach the round-of-sixteen (88.4% and 86.5%, respectively), while the probabilities to at least reach the quarter finals differ significantly. While Spain goes at least to the quarter finals with a probability of 73.1%, Germany only achieves a probability of 58.0%. Obviously, in contrast to Spain, Germany has a rather high chance to meet a strong opponent in the round-of-sixteen. In case they reach the round-of-sixteen, Germany would face Brazil, Switzerland, Serbia or Costa Rica, while Spain would face Uruguay, Russia, Saudi Arabia or Egypt. In the following rounds, Germany starts catching up to Spain finally ending up with almost equal winning probabilities.

---

[5]The final group standings are determined by (1) the number of points, (2) the goal difference and (3) the number of scored goals. If several teams coincide with respect to all of these three criteria, a separate chart is calculated based on the matches between the coinciding teams only. Here, again the final standing of the teams is determined following criteria (1)-(3). If still no distinct decision can be taken, the decision is taken by lot.

Table 8: Estimated probabilities (in %) for reaching the different stages in the FIFA World Cup 2018 for all 32 teams based on 100,000 simulation runs of the FIFA World Cup together with winning probabilities based on the ODDSET odds.

|  |  |  | Round of 16 | Quarter finals | Semi finals | Final | World Champion | Oddset |
|---|---|---|---|---|---|---|---|---|
| 1. |  | ESP | 88.4 | 73.1 | 47.9 | 28.9 | 17.8 | 11.8 |
| 2. |  | GER | 86.5 | 58.0 | 39.8 | 26.3 | 17.1 | 15.0 |
| 3. |  | BRA | 83.5 | 51.6 | 34.1 | 21.9 | 12.3 | 15.0 |
| 4. |  | FRA | 85.5 | 56.1 | 36.9 | 20.8 | 11.2 | 11.8 |
| 5. |  | BEL | 86.3 | 64.5 | 35.7 | 20.4 | 10.4 | 8.3 |
| 6. |  | ARG | 81.6 | 50.5 | 29.8 | 15.2 | 7.3 | 8.3 |
| 7. |  | ENG | 79.8 | 57.0 | 29.8 | 15.6 | 7.1 | 4.6 |
| 8. |  | POR | 67.5 | 46.1 | 19.8 | 7.3 | 2.5 | 3.8 |
| 9. |  | CRO | 65.9 | 30.8 | 15.6 | 6.0 | 2.2 | 3.0 |
| 10. |  | SUI | 58.9 | 30.6 | 13.1 | 5.6 | 2.2 | 1.0 |
| 11. |  | COL | 79.2 | 33.1 | 14.0 | 5.7 | 2.1 | 1.8 |
| 12. |  | DEN | 59.0 | 26.1 | 12.4 | 4.8 | 1.7 | 1.1 |
| 13. |  | URU | 86.6 | 37.5 | 13.5 | 4.4 | 1.3 | 2.8 |
| 14. |  | SWE | 54.0 | 21.7 | 8.0 | 3.1 | 1.0 | 0.8 |
| 15. |  | POL | 60.6 | 18.9 | 6.8 | 2.3 | 0.7 | 1.5 |
| 16. |  | PER | 39.2 | 15.4 | 6.6 | 2.1 | 0.6 | 0.4 |
| 17. |  | ICE | 36.6 | 12.9 | 5.3 | 1.7 | 0.5 | 0.6 |
| 18. |  | SRB | 36.2 | 13.8 | 4.7 | 1.5 | 0.4 | 0.6 |
| 19. |  | SEN | 39.7 | 10.9 | 3.7 | 1.1 | 0.3 | 0.6 |
| 20. |  | MOR | 30.3 | 14.8 | 4.0 | 1.0 | 0.3 | 0.3 |
| 21. |  | TUN | 22.8 | 8.9 | 2.8 | 0.8 | 0.2 | 0.2 |
| 22. |  | MEX | 41.5 | 13.9 | 3.7 | 1.1 | 0.2 | 1.0 |
| 23. |  | CRC | 21.4 | 6.4 | 1.7 | 0.4 | 0.1 | 0.3 |
| 24. |  | EGY | 45.5 | 10.3 | 2.1 | 0.4 | 0.1 | 0.6 |
| 25. |  | RUS | 50.4 | 10.5 | 2.4 | 0.4 | 0.1 | 2.2 |
| 26. |  | NGA | 15.8 | 4.0 | 1.2 | 0.3 | 0.1 | 0.6 |
| 27. |  | AUS | 16.2 | 4.2 | 1.2 | 0.3 | 0.1 | 0.3 |
| 28. |  | JPN | 20.5 | 4.1 | 0.9 | 0.2 | 0.0 | 0.6 |
| 29. |  | KOR | 17.9 | 4.0 | 0.8 | 0.2 | 0.0 | 0.6 |
| 30. |  | IRN | 13.8 | 5.1 | 0.9 | 0.1 | 0.0 | 0.3 |
| 31. |  | PAN | 11.1 | 2.5 | 0.5 | 0.1 | 0.0 | 0.1 |
| 32. |  | KSA | 17.5 | 2.6 | 0.4 | 0.0 | 0.0 | 0.1 |

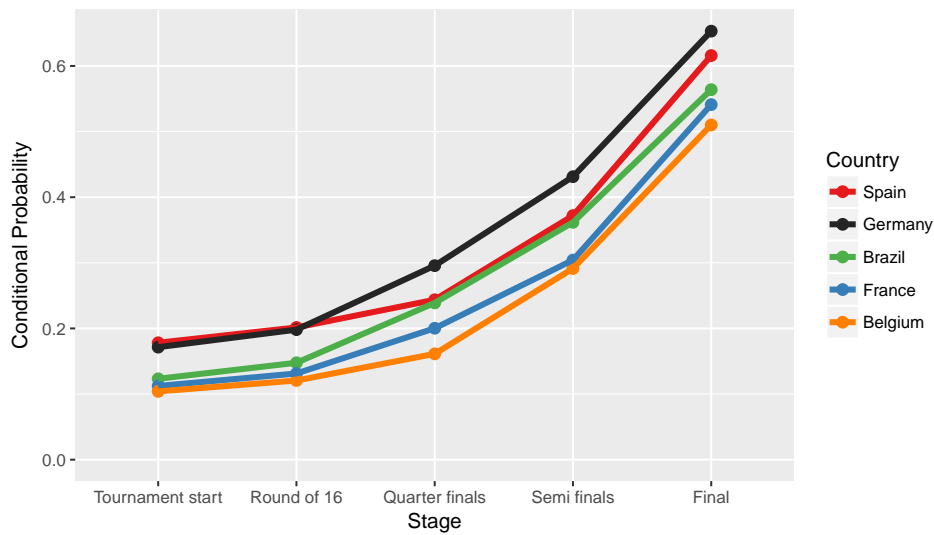

Figure 4: Winning probabilities conditional on reaching the single stages of the tournament for the five favored teams.

Figure 4 further illustrates this influence of the tournament design on the winning probabilities. It shows the winning probabilities conditional on reaching the single stages of the tournament for the five favored teams. All teams start with the probabilities displayed in Table 8 and, accordingly, their respective (conditional) probabilites increase with each stage. Again, the comparison between Spain and Germany is interesting. The fact that overall Spain is slightly favored over Germany is mainly due to the fact that Germany has a comparatively high chance to drop out in the round-of-sixteen. Conditional on reaching the quarter finals, Germany overtakes Spain and is (from this tournament stage on) the favored team.

## 4.2 Most probable tournament course

Finally, based on the 100,000 simulations, we also provide the most probable tournament course. Here, for each of the eight groups we selected the most probable final group standing, also considering the order of the first two places, but without considering the irrelevant order of the teams on places three and four. The results together with the corresponding probabilities are presented in Table 9.

Obviously, there are large differences with respect to the groups' balances. While in Group B and Group G the model forecasts Spain followed by Portugal as well as Belgium followed by England with rather high probabilities of 38.5% and 38.1%, respectively, other groups such as Group A, Group F and Group H seem to be more volatile.

Moreover, we provide the most probable course of the knockout stage in Figure 5. The most likely round-of-sixteen directly results from those teams qualifying for the knockout stage in Table 9. For all following matches we compute the probabilities for the respective two teams (say team A and team B) to go to the next stage. This is done by applying the Skellam distribution to first get the probabilities for *A wins*, *draw* and *B wins* after 90 minutes. Second, the probability for *draw* is distributed between teams A and B again following the principles of extra-time and penalty shootouts we already apply for draws in the knockout stage in the previous section. This way, finally the probabilities for *A wins* and *B wins* add up to 1, as it is necessary for the knockout stage. In Figure 5, the probabilities accompanying the edges of the tournament tree represent the probability of the favored team to proceed to the next stage.

According to the most probable tournament course, instead of the Spanish the German team would win the World Cup. However, again it becomes obvious that with (in that case) Switzerland the German team has to face a much stronger opponent than Spain in the round-of-sixteen. Even though still being the favorite in this match, they would succeed to move on to the quarter finals only with a probability of 61%. While in the most probable course of the knock-out stage, though having tough times in all single stages, Germany would make its way into the final and defend the title, the previous section showed that generally still Spain is the most likely winner.

We wish to attract the reader's attention to the fact that, despite being the most probable tournament course, due to the myriad of possible constellations this exact tournament course is still extremely unlikely: if we take the product of all single probabilities of Table 9 and Figure 5, its overall probability yields $1.55 \cdot 10^{-5}\%$. Hence, deviations of the true tournament course from the model's most probable one are not only possible, but very likely.

Table 9: Most probable final group standings together with the corresponding probabilities for the FIFA World Cup 2018 based on 100,000 simulation runs.

| Group A 28.7% | Group B 38.5% | Group C 31.5% | Group D 30.7% |
|---|---|---|---|
| 1. URU | 1. ESP | 1. FRA | 1. ARG |
| 2. RUS | 2. POR | 2. DEN | 2. CRO |
| KSA | MOR | AUS | ICE |
| EGY | IRN | PER | NGA |

| Group E 29.0% | Group F 29.9% | Group G 38.1% | Group H 26.5% |
|---|---|---|---|
| 1. BRA | 1. GER | 1. BEL | 1. COL |
| 2. SUI | 2. SWE | 2. ENG | 2. POL |
| CRC | MEX | PAN | SEN |
| SRB | KOR | TUN | JPN |

## 5 Concluding remarks

In this work, we first compared three different modeling approaches for the scores of soccer matches with regard to their predictive performances based on all matches from the four previous FIFA World Cups 2002 – 2014, namely *random forests*, *Poisson regression models* and *ranking methods*. The former two approaches incorporate covariate information of the opposing teams, while the latter method pro-vides team ability parameters which serve as adequate estimates of the current team strengths. The comparison revealed that the best-performing prediction methods on the training data were the ranking methods and the random forests. However, we then showed that by incorporating the team ability parameters from the ranking methods as an additional covariate into the random forest the predictive power becomes substantially increased, leading to the best model capable of beating the bookmakers.

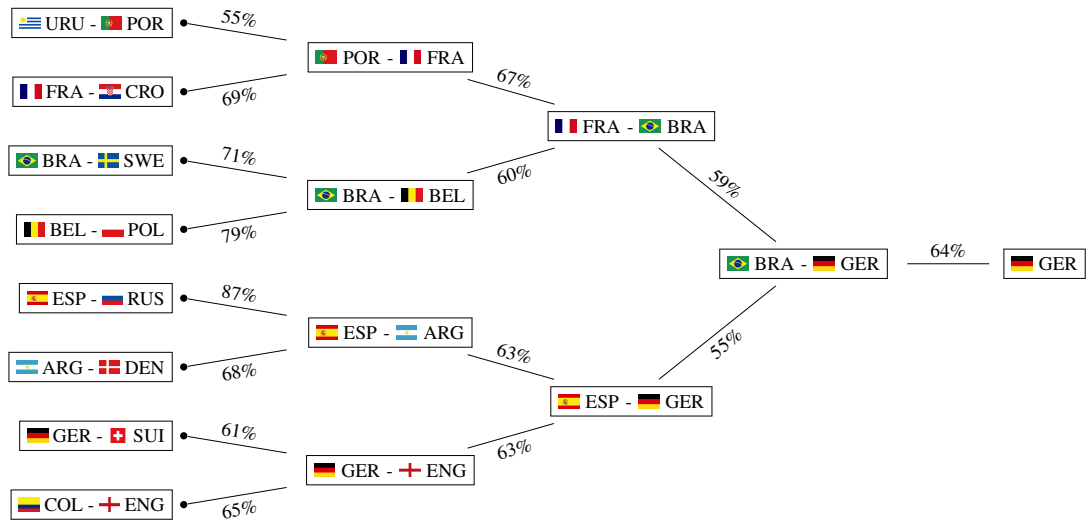

Figure 5: Most probable course of the knockout stage together with corresponding probabilities for the FIFA World Cup 2018 based on 100,000 simulation runs.

We chose this random forest method as the most promising candidate and fitted it to a training data set containing all matches of the four previous FIFA World Cups 2002 – 2014. Based on the corresponding estimates, we repeatedly simulated the FIFA World Cup 2018 100,000 times. According to these simulations, Spain and Germany turned out to be the top favorites for winning the title, with a slight advantage for Spain. Furthermore, survival probabilities for all teams and at all tournament stages as well as the most probable tournament course are provided. Interestingly, regarding the most probable tournament course, Germany would take home the trophy.

As also the bookmakers list the German team as the top favorite, the model's forecast of Spain as the most likely tournament winner might be surprising at first glance. Hence, it is worth to have a deeper look into the single tournament stages.

By analyzing the winning probabilities conditional on reaching the single stages of the tournament it turns out that the fact that overall Spain is slightly favored over Germany is mainly due to the fact that Germany has a comparatively high chance to drop out in the round-of-sixteen. Actually, conditioned that Germany reaches the quarter finals, it overtakes Spain and is (from this tournament stage on) the favored team.

# References

Audran, J., M. Bolliger, T. Kolb, J. Mariscal, and Q. Pilloud (2018): "Investing and football - Special edition: 2018 World Cup in Russia," Working paper, UBS.

Boshnakov, G., T. Kharrat, and I. G. McHale (2017): "A bivariate weibull count model for forecasting association football scores," *International Journal of Forecasting*, 33, 458 – 466, URL http://www.sciencedirect.com/science/article/pii/S0169207017300018.

Breiman, L. (2001): "Random forests," *Machine Learning*, 45, 5–32.

Breiman, L., J. H. Friedman, R. A. Olshen, and J. C. Stone (1984): *Classification and Regression Trees*, Monterey, CA: Wadsworth.

Dixon, M. J. and S. G. Coles (1997): "Modelling association football scores and inefficiencies in the football betting market," *Journal of the Royal Statistical Society: Series C (Applied Statistics)*, 46, 265–280.

Dyte, D. and S. R. Clarke (2000): "A ratings based Poisson model for World Cup soccer simulation," *Journal of the Operational Research Society*, 51 (8), 993–998.

Friedman, J., T. Hastie, and R. Tibshirani (2010): "Regularization paths for generalized linear models via coordinate descent," *Journal of Statistical Software*, 33, 1.

Gneiting, T. and A. Raftery (2007): "Strictly proper scoring rules, prediction, and estimation," *Journal of the American Statistical Association*, 102, 359–376.

Groll, A. and J. Abedieh (2013): "Spain retains its title and sets a new record - generalized linear mixed models on European football championships," *Journal of Quantitative Analysis in Sports*, 9, 51–66.

Groll, A., T. Kneib, A. Mayr, and G. Schauberger (2018): "On the dependency of soccer scores – A sparse bivariate Poisson model for the UEFA European Football Championship 2016," *Statistical Modelling*, to appear.

Groll, A., G. Schauberger, and G. Tutz (2015): "Prediction of major international soccer tournaments based on team-specific regularized Poisson regression: an application to the FIFA World Cup 2014," *Journal of Quantitative Analysis in Sports*, 11, 97–115.

Hoerl, A. E. and R. W. Kennard (1970): "Ridge regression: Biased estimation for nonorthogonal problems," *Technometrics*, 12, 55–67.

Hothorn, T., P. Buehlmann, T. Kneib, M. Schmid, and B. Hofner (2017): *mboost: Model-Based Boosting*, URL https://CRAN.R-project.org/package=mboost, R package version 2.8-1.

Hothorn, T., P. Bühlmann, S. Dudoit, A. Molinaro, and M. J. van der Laan (2006): "Survival ensembles," *Biostatistics*, 7, 355–373.

Hothorn, T., B. Lausen, A. Benner, and M. Radespiel-Tröger (2004): "Bagging survival trees," *Statistics in Medicine*, 23, 77–91.

Karlis, D. and I. Ntzoufras (2003): "Analysis of sports data by using bivariate poisson models," *The Statistician*, 52, 381–393.

Koopman, S. J. and R. Lit (2015): "A dynamic bivariate poisson model for analysing and forecasting match results in the english premier league," *Journal of the Royal Statistical Society: Series A (Statistics in Society)*, 178, 167–186.

Leitner, C., A. Zeileis, and K. Hornik (2010a): "Forecasting sports tournaments by ratings of (prob)abilities: A comparison for the EURO 2008," *International Journal of Forecasting*, 26 (3), 471–481.

Leitner, C., A. Zeileis, and K. Hornik (2010b): "Forecasting the winner of the FIFA World Cup 2010," Research Report Series Report 100, Department of Statistics and Mathematics, University of Vienna.

Ley, C., T. Van de Wiele, and H. Van Eetvelde (2018): "Ranking soccer teams on basis of their current strength: a comparison of maximum likelihood approaches," *Statistical Modelling*, submitted.

Maher, M. J. (1982): "Modelling association football scores," *Statistica Neerlandica*, 36, 109–118.

McHale, I. and P. Scarf (2007): "Modelling soccer matches using bivariate discrete distributions with general dependence structure," *Statistica Neerlandica*, 61, 432–445, URL https://onlinelibrary.wiley.com/doi/abs/10.1111/j.1467-9574.2007.00368.x.

McHale, I. G. and P. A. Scarf (2011): "Modelling the dependence of goals scored by opposing teams in international soccer matches," *Statistical Modelling*, 41, 219–236.

Quinlan, J. R. (1986): "Induction of decision trees," *Machine learning*, 1, 81–106.

R Core Team (2018): *R: A Language and Environment for Statistical Computing*, R Foundation for Statistical Computing, Vienna, Austria, URL https://www.R-project.org/.

Schauberger, G. and A. Groll (2018): "Predicting matches in international football tournaments with random forests," *Statistical Modelling*, in press.

Strobl, C., A.-L. Boulesteix, T. Kneib, T. Augustin, and A. Zeileis (2008): "Conditional variable importance for random forests," *BMC Bioinformatics*, 9, 307.

Strobl, C., A.-L. Boulesteix, A. Zeileis, and T. Hothorn (2007): "Bias in random forest variable importance measures: Illustrations, sources and a solution," *BMC Bioinformatics*, 8, 25.

Tibshirani, R. (1996): "Regression shrinkage and selection via the lasso," *Journal of the Royal Statistical Society*, B 58, 267–288.

Wright, M. N. and A. Ziegler (2017): "ranger: A fast implementation of random forests for high dimensional data in C++ and R," *Journal of Statistical Software*, 77, 1–17.

Yuan, M. and Y. Lin (2006): "Model selection and estimation in regression with grouped variables," *Journal of the Royal Statistical Society*, B 68, 49–67.

Zeileis, A., C. Leitner, and K. Hornik (2012): "History repeating: Spain beats Germany in the EURO 2012 final," Working paper, Faculty of Economics and Statistics, University of Innsbruck.

Zeileis, A., C. Leitner, and K. Hornik (2014): "Home Victory for Brazil in the 2014 FIFA World Cup," Working paper, Faculty of Economics and Statistics, University of Innsbruck.

Zeileis, A., C. Leitner, and K. Hornik (2016): "Predictive Bookmaker Consensus Model for the UEFA Euro 2016," Working Papers 2016-15, Faculty of Economics and Statistics, University of Innsbruck, URL http://EconPapers.repec.org/RePEc:inn:wpaper:2016-15.

Zeileis, A., C. Leitner, and K. Hornik (2018): "Probabilistic forecasts for the 2018 FIFA World Cup based on the bookmaker consensus model," Working Paper 2018-09, Working Papers in Economics and Statistics, Research Platform Empirical and Experimental Economics, Universität Innsbruck, URL http://EconPapers.RePEc.org/RePEc:inn:wpaper:2018-09.